\documentclass[10pt,journal]{IEEEtran}
\usepackage{amsmath,amsthm,amsfonts,amssymb,bm,mathrsfs}
\usepackage{graphicx,subfigure}
\usepackage{epstopdf}
\usepackage{cite}
\usepackage{url}
\usepackage{array}
\usepackage{multirow,booktabs,threeparttable}
\usepackage{algorithm}
\usepackage{algorithmic}
\usepackage{color}
\usepackage{tabularx}
\usepackage[square, comma, sort&compress, numbers]{natbib}

\def\BibTeX{{\rm B\kern-.05em{\sc i\kern-.025em b}\kern-.08em
		T\kern-.1667em\lower.7ex\hbox{E}\kern-.125emX}}

\graphicspath{{fig/}}

\begin{document}
\title{Graph Attention Network-based Multi-agent Reinforcement Learning for Slicing Resource Management in Dense Cellular Network}
\author{\IEEEauthorblockN{Yan Shao, Rongpeng Li, Bing Hu, Yingxiao Wu, Zhifeng Zhao and Honggang Zhang}
		\thanks{
			\copyright 2015 IEEE. Personal use of this material is permitted. However, permission to use this material for any other purposes must be obtained from the IEEE by sending a request to pubs-permissions@ieee.org.
			
			This work was supported in part by National Key R\&D Program of China (No. 2020YFB1804804), National Natural Science Foundation of China (No. 61731002, 62071425), Zhejiang Key Research and Development Plan (No. 2019C01002, 2019C03131), Huawei Cooperation Project, the Project sponsored by Zhejiang Lab (No. 2019LC0AB01), and Zhejiang Provincial Natural Science Foundation of China (No. LY20F010016). \textit{(Corresponding author: Rongpeng Li.)}
			
			Y. Shao, R. Li, B. Hu and H. Zhang are with the College of Information Science and Electronic Engineering, Zhejiang University, Hangzhou 310027, China (e-mail: \{shaoy, lirongpeng, binghu, honggangzhang\}@zju.edu.cn).
			
			Y. Wu is with the College of Computer Science \& Technology, Hangzhou Dianzi University, Hangzhou 310027, China. (e-mail: wuyx@zhejianglab.com).
			
			Z. Zhao is with Zhejiang Lab, Hangzhou, China as well as the College of Information Science and Electronic Engineering, Zhejiang University, Hangzhou 310027, China (e-mail: zhaozf@zhejianglab.com).
			
			Part of the paper has been accepted by IEEE WCNC 2021 \cite{shao2021graph}.
			
		}}

\maketitle

\begin{abstract}
	Network slicing (NS) management devotes to providing various services to meet distinct requirements over the same physical communication infrastructure and allocating resources on demands. Considering a dense cellular network scenario that contains several NS over multiple base stations (BSs), it remains challenging to design a proper real-time inter-slice resource management strategy, so as to cope with frequent BS handover and satisfy the fluctuations of distinct service requirements. In this paper, we propose to formulate this challenge as a multi-agent reinforcement learning (MARL) problem in which each BS represents an agent. Then, we leverage graph attention network (GAT) to strengthen the temporal and spatial cooperation between agents. Furthermore, we incorporate GAT into deep reinforcement learning (DRL) and correspondingly design an intelligent real-time inter-slice resource management strategy. More specially, we testify the universal effectiveness of GAT for advancing DRL in the multi-agent system, by applying GAT on the top of both the value-based method deep Q-network (DQN) and a combination of policy-based and value-based method advantage actor-critic (A2C). Finally, we verify the superiority of the GAT-based MARL algorithms through extensive simulations.
\end{abstract}


\section{Introduction}
	
	The fifth-generation (5G) mobile system devotes to offering supports for tremendous subscribers with diverse service requirements \cite{li2018deep}. A total of 190 million 5G subscribers are expected by the end of 2020. In 2025, 5G networks will carry nearly 45 percent of the world mobile data traffic and cover up to 65 percent of the demands of global population \cite{eric2020}. The large amount and sharp growth of data traffic has brought severe pressure to current mobile networks, which gives rise to the research, aiming at the improvements of the network throughput, utilization, quality of service (QoS), and the combinations thereof. Facing such huge traffic demands, current researches mainly focus on two schemes which complement each other based on 5G. The evolutionary scheme aims to scale up and improve the efficiency of mobile networks including but not limited to spectrum reuse, massive multiple-input and multiple-output (MIMO) and higher frequency bands (e.g., millimeter-wave and Tera-Hertz communications) \cite{foukas2017network}. The other one is service-oriented trying to cater to a wide range of services differing in their requirements and types of devices which is also the focus of this article. Three typical scenarios serving for diverse demands based on this scheme are enhanced mobile broadband (eMBB), massive machine-type communications (mMTC), and ultra-reliable and low-latency communications (URLLC). The stack differences of these scenarios are three folds: (a) eMBB provides subscribers with stable and high peak data rates to cater the typical services like 4k/8k HD, AR/VR, holographic image, etc; (b) mMTC commits to supporting the massive Internet of Things (IoT) devices, which need no excessive data payloads; (c) URLLC furnishes with ultra-reliability and low-latency which meets the industrial requirements such as automatic driving, telemedicine and so on \cite{popovski20185g}. These differentiated vertical services bring pressures for mobile operators. Hence service-oriented scheme requires a radical rethink of 5G mobile system and its infrastructure to turn into the more flexible and programmable fabric
	
	As a non-nascent concept, network slicing (NS), which benefits from the advances of software defined networking (SDN) and network functions virtualization (NFV), has been proposed to facilitate the customized end-to-end network services to help operators launch resource with more flexibility and cost-efficiency to market. In other words, \cite{zhou2016network} puts forward that NS could act as a service (NSaaS). As an end-to-end service, NS has been proposed for core networks (CN) initially. After that, the Third Generation Partnership Project (3GPP) considers that radio access networks (RAN) also need specific functionalities to support multiple slices or even partition resources for different NS \cite{da2016impact,3gpp2016study,zhou2016network}. Thus, 5G system becomes capable to provide customized end-to-end network slices from CN to RAN. Similar to traditional resource allocation schemes, NS allows different tenants to share the same communications and computing resources. However, NS involves more complicated factors, as it aims to provide the dedicated fully-functional virtual network according to diverse requirements such as ultra-low latency in URLLC, ultra-high throughput in eMBB, and other customized services. Each virtual network is allocated a certain amount of resources and then re-allocates them to its subscribers based on specific rules. In this regard, NS implies allocating resources in a multiple-tier manner, and each tier has different constraints. In this way, the physical and computational resources are relatively more flexible and independent with slight interference than single network resources. To achieve the vision of NS and provide a smoother network experience for subscribers, the mainstream research contents mainly focus on intra-slice spectrum reuse, efficient inter-slice handover mechanism \cite{sun2020efficient}, dynamic inter-slice resource management, etc.
	
	A proper real-time inter-slice resource management strategy can promote network performance by meeting distinct service requirements and relieve the pressure caused by volatile demand variations while maintaining acceptable spectrum efficiency (SE). But the fluctuations of service demands in RAN is very unstable while the mobility of subscribers intensifies these fluctuations. These factors result in the failure of classical dedicated resource management strategies which lack the flexibility to change their strategies in real time. Recently, some researchers propose to use reinforcement learning (RL) to fix out this problem such as deep Q-network (DQN) \cite{li2018deep}, generative adversarial network-powered deep distributional Q network (GAN-DDQN) \cite{hua2019gan} and long short-term memory-based advantage actor-critic (LSTM-A2C) \cite{li2020lstm}. However, these works mainly consider single base station (BS) scenarios and ignore the significance of cooperation among BSs. In fact, RAN in 5G mobile system is conceived as a dense cellular network due to the adoption of higher frequency bands and the incident smaller coverage. Thus, strengthening the cooperation and obtaining the related information from adjacent BSs is helpful to design an efficient resource management strategy for the current BS. Intriguingly, Graph Attention Network (GAT) \cite{velivckovic2017graph} is such an effective way to address the cooperation issue by processing structured data from multiple BSs as a graph. Accordingly, this paper primarily considers a dense network scenario with moving subscribers in which each BS is regraded as an agent, and proposes a multi-agent reinforcement learning (MARL) algorithm which combines Graph Attention Network (GAT) with two types of typical DRL algorithms (i.e., DQN and its variants, as well as A2C) to provide more precise resource management strategies. The main contributions are as follows: 
	\begin{itemize}
		\item We build up a cooperation mechanism among BSs through GAT to capture and process the pattern of fluctuant service demands in temporal and spatial domains. We construct the multi-BS scenario as an adjacent graph and define the neighborhood by Euclidean distance. On this basis, we leverage GAT to aggregate the information from adjacent BSs that achieve dynamic collaboration among BSs in real-time. Moreover, we involve multiple GAT layers which can expand receptive field under the same communication conditions. 
		\item We propose a succinct and universal reward function to replace those complex clipping and shaping functions. It only has several hyper-parameters related to the optimization objective to be adjusted and can be interpreted easily. 
		\item We employ mainstream RL algorithms \cite{hessel2017rainbow} to optimize the real-time inter-slice resource management strategy among various NS. In particular, we use a value-based RL, DQN and its variants (i.e., double DQN and dueling DQN), to forecast the actions of resource management more precisely. Besides using the value-based RL method, we proactively involve a combination algorithm of policy-based and value-based method, A2C, to obtain an optimal policy for resource management. 
		Applying GAT to different RL algorithms effectively demonstrates the universality of GAT in promoting the performance of MARL algorithms in multi-agent systems.
		\item We verify the performance of GAT-based MARL algorithms in the simulation containing subscribers with various trajectories in temporal and spatial domains, which is more realistic and adds to the difficulty of predictions. Besides, We compare the GAT-based MARL algorithms to normal algorithms and verify the superiority of our work.
	\end{itemize}
	
	The reminder of the paper is organized as follows: Section II overview the related work. Section III presents the system model and formulates this problem as a Markov Game (MG) which can be fixed out in MARL algorithms. The details of GAT-based MARL algorithms for resource management are illustrated in Section IV. Then, we provide the numerical analysis and simulation results in Section V. In the end, Section VI summarizes the above works and gives future research directions.
	 

\section{Related Work}
	When addressing the real-time resource management among diverse NS, the utility of RAN resources is supposed to be maximized for the better-performing and cost-efficient services. Referring to \cite{zhou2016network,li2018deep,hua2019gan,li2020lstm}, the utility performance in RAN is generally measured by (a) SE since spectrum resource is scarce in RAN; (b) the service level agreement (SLA) satisfaction ratio (SSR) within the slice tenants, which usually imposes stringent requirement and reflects the QoS perceived by subscribers. 
	
	From the viewpoint of resource utilization, spectrum reuse alleviates the problem of resource scarcity in RAN through opportunistic spectrum access (OSA) \cite{huang2008osa}. OSA allows secondary subscribers to identify and exploit the unused spectrum owned by primary subscribers opportunistically while limiting the interference to primary subscribers below a predefined threshold. However, OSA cannot ensure the quality of services to secondary subscribers. In particular, we mainly focus on the real-time inter-slice resource management among various network slices which are exclusive of tenants to satisfy the customized services with specific requirements such as ultra-low latency in URLLC, ultra-high throughput in eMBB, and other customized services. In other words, rigorous requirement should be satisfied for all subscribers. Thus, it is not suitable to directly apply the spectrum reuse method to NS, and efforts has to be taken, so as to make the transmission more better-performing and cost-efficient. 
	
	\begin{figure*}[htbp]
		\centering
		\includegraphics[width=0.7\textwidth]{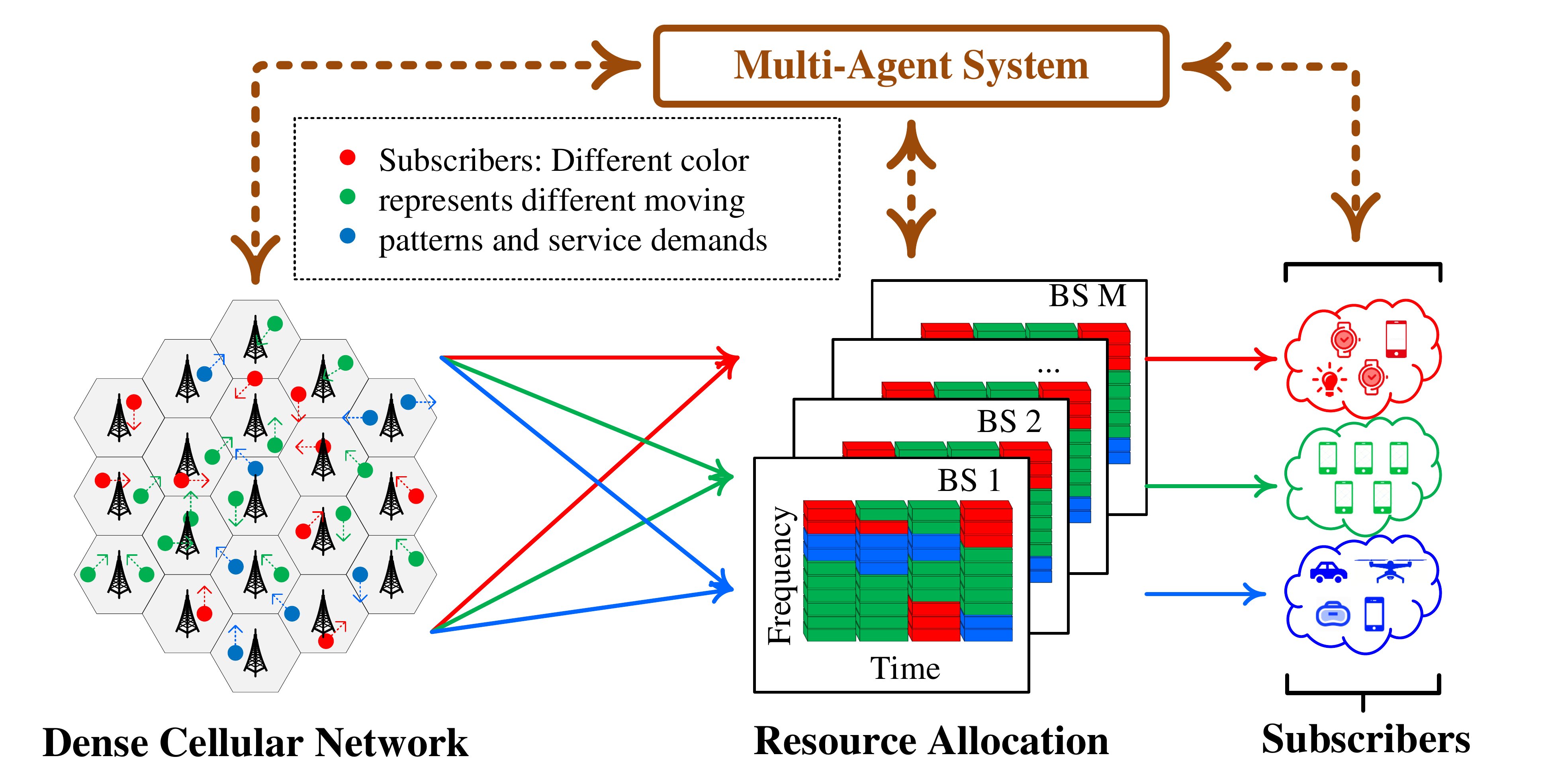}
		\caption{The radio access network scenario with multiple base stations and a number of moving subscribers.}
		\label{fig:system_model}
	\end{figure*}
	
	Moreover, the actual demands of each NS are not only diverse but also dynamic due to the mobility of subscribers and requirement variations. It fails the classical dedicated resource management strategies which lack flexibility and the ability to change their strategies in real time. Hitherto, other meaningful solutions have been presented. \cite{han2017modeling} proposes a profit optimization model with a value chain to analyze the profit of each slice and optimize the strategy based on the traditional business mode. However, it requires that tenants have a priori knowledge about the service demands and the cost/revenue models of every slice which seems insatiable. Subsequently, the authors put forward an online genetic approach by encoding feasible slicing strategies into an individual binary sequence \cite{han2018slice} but not considering the influence of various service requirements and SLA in each slice. \cite{vo2018slicing} considers the radio bandwidth, caching, and backhaul components jointly, and models the resource management as a bi-convex problem which would be solved by numerical solutions. But this optimization problem is intractable when the parameters are scaled up for the increasing of NS or the shareable resources. \cite{sun2019user} mainly focuses on access control and resource management of NS for a scenario with multiple BSs. But, it impractically assumes that the demand rate is fixed for every subscriber. Thus, despite the satisfactory numerical simulation results given by the above works, it involves some impractical assumptions and becomes difficult to directly apply the optimization tools or heuristic algorithms backed by complex numerical analysis in resource management, due to the lack of flexibility and extensibility. For example, when the scenario parameters are changed such as requesting more stringent SLA, facing moving subscribers and adjusting the shared resources, these methods may no longer be applicable.
	
	Given the well-known success of AlphaGo \cite{silver2016mastering}, deep reinforcement learning (DRL) comes to the attention of the public. DRL focuses on promoting agents to learn an optimal policy by interacting with the environment and reinforcing the tendency policy producing higher rewarding consequences \cite{sutton2018reinforcement}. This characteristic makes it outstanding in many fields such as power control \cite{nasir2018deep}, green communications \cite{liu2018deepnap}, cloud radio access networks \cite{xu2017deep} and mobile edge computing and caching \cite{he2017software}. Considering this powerful ability, some researchers tend to leverage DRL to address the real-time resource management in RAN. The previous work in \cite{li2018deep} firstly uses DQN, a typical type of DRL, to match the allocated resource to multiple slices based on the fluctuant demands of subscribers. It verifies that DQN could obtain the deep relationship between the demands of subscribers and allocated resources in resource-constrained scenarios. Based on this work, the effects of random noise on the calculation of SE and SSR are further studied in \cite{han2018slice}. They propose GAN-DDQN to learn the action-value distribution driven by minimizing the discrepancy between the estimated action-value distribution and the target action-value distribution. Furthermore, \cite{li2020lstm} intends to incorporate the LSTM into A2C to track the temporal patterns of demands caused by the mobility of subscribers and thus improves the system utility. 
	
	However, the aforementioned methods mainly do not take the significance of cooperation among BSs into consideration. Strengthening the cooperation can capture the moving trajectories of subscribers for catering to the temporal and spatial fluctuations of service demands and boost the learning efficiency, which is meaningful in the dense cellular network of 5G.  Therefore, we propose a GAT-based MARL algorithm to provide more precise resource management strategies.

\section{System model and problem definition}

\subsection{System Model}
			
	
	In this section, we design a multi-agent system model which simulates a RAN scenario synthetically consisting of multiple BSs and moving subscribers as depicted in Fig.~\ref{fig:system_model}. The main purpose of this paper is to optimize the inter-slice resource management strategy for each BS in real time according to the various demands of subscribers when primarily considering the downlink transmissions only. Different from the previous works in \cite{hua2019gan,li2020lstm}, a more practical scenario with multiple BSs and several subscribers with intricate mobility patterns is taken into consideration. Without loss of generality, this scenario is conceived to be a dense cellular network with $M$ BSs. The set of BSs is represented by $\mathcal{B}$. The assigned bandwidth for each BS is $W$, which is shared by $N$ NS, expressed by $\mathcal{N}, |\mathcal{N}|=N$.  The set of subscribers is represented by $\mathcal{U}$. We use $\mathcal{U}_{mn}$ to denote the set of subscribers which demand the services provided by $n^{th}$ NS in the $m^{th}$ BS.
	
	We conceive that the inter-slice resource management strategy is updated in a timeslot model corresponding to the demands of subscribers periodically. The fluctuant demands for diverse NS in the $m^{th}$ BS are $\bm{d}_m=\{d_{m1},\dots,d_{mn},\dots,d_{mN}\}$, the determinant factor for the resource management strategy of BSs. We use $\bm{w}_m=\{w_{m1},\dots,w_{mn},\dots,w_{mN}\}$ to represent the inter-slice resource management strategy for the $m^{th}$ BS.
	
	To achieve the aforementioned objective, (i.e., optimizing the inter-slice resource management strategy), the system utility $J$ is introduced as a vital evaluation criterion, composed by the weighted sum of SE and SSR. We can formulate this optimization as follows:
	\begin{equation}
		\label{optimization}
		\begin{aligned}
			\mathop{\max}_{\bm{w}_m} \quad & J_m= \alpha \cdot \mathrm{SE}_m (\bm{d_m},\bm{w_m})\\
			&\qquad + \sum_{n \in N} \beta_{n} \cdot \mathrm{SSR}_{mn} (\bm{d_m},\bm{w_m}) \\
			s.t.\quad &  \sum_{n=1}^{N} w_{mn}= W  \\
			& w_{mn}=c \cdot \Delta, \ \forall n \in [1,\cdots,N]
		\end{aligned}
	\end{equation}
	where $\Delta$ is the minimum allocated bandwidth granularity for per slice based on the size of resource block which means the bandwidth allocated for per slice is several times of $\Delta$ while the magnification is determined by an integer $c$. $\alpha$ and $\bm{\beta}=\{ \beta_1, \dots, \beta_{N} \}$ are the hyper-parameters of the weighted sum representing the relative importance of SE and SSR which can be set according to the practical system requirements. We also test different combinations of $\alpha$ and $\bm{\beta}$ in Section V.B. Intuitively, larger $\bm{\beta} $ implies stronger emphasis on satisfying SLA but might degrade the SE so that we need to trade off between SSR and SE.
	
	Thereinto, SE could be obtained from the downlink signal-to-noise ratio (SNR) according to the Shannon capacity. We define that $r_{u_{mn}}$ represents the downlink data rate of subscriber $u_{mn}$ served by $ n^{th}$ NS in $m^{th}$ BS. For simplicity, it is described as
	
	\begin{equation}
		r_{u_{mn}} = w_{mn} \log (1+\mathrm{SNR}_{u_{mn}}), \forall u_{mn} \in \mathcal{U}_{mn}
	\end{equation}
	where $ \mathrm{SNR}_{u_{mn}} $ is the downlink signal-to-noise ratio between subscriber $u_{mn}$ and $m^{th} $ BS, defined as:
	\begin{equation}
		\mathrm{SNR}_{u_{mn}}=\frac{g_{u_{mn}} P_{u_{mn}}}{N_0 w_{mn}}
	\end{equation}
	where $g_{u_{mn}}$ is the average channel gain composed by the path loss and shadowing which are decided by the channel model, $P_{u_{mn}}$ is the transmission power, and $N_0$ is the single-side noise spectral density.
	Next, SE can be calculated by:
	\begin{equation}
		\mathrm{SE}_m = \frac{\sum_{n \in N} \sum_{u_{mn} \in \mathcal{U}_{mn}} r_{u_{mn}}}{W}
	\end{equation}
	Due to the bandwidth limitation in Eq.~(\ref{optimization}), $\sum_{n=1}^{N} w_{mn}= W$, thus the scale of SE is decided by the SNR of channel mode. Moreover, downlink data rate is a significant component of SE which means higher data rate leads to higher SE.
	
	\begin{figure}[tbp]
		\centering
		\subfigure[iteration:290]{
			\centering
			\includegraphics[width=0.475\textwidth]{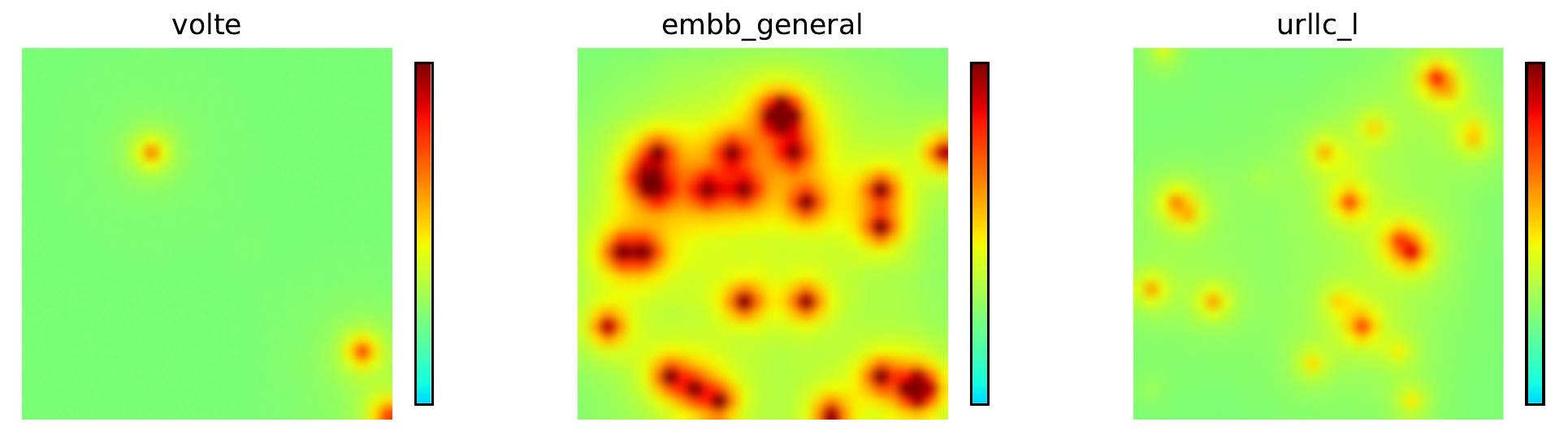}
			\label{fig:290}
		}
		\subfigure[iteration:300]{
			\centering
			\includegraphics[width=0.475\textwidth]{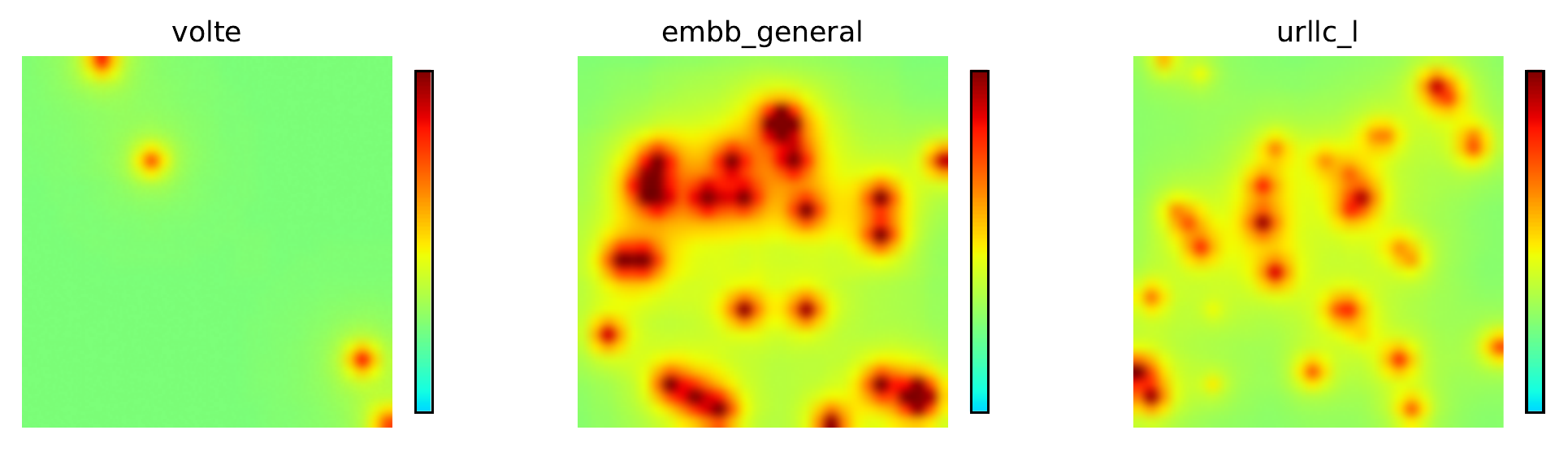}
			\label{fig:300}
		}
		\subfigure[iteration:310]{
			\centering
			\includegraphics[width=0.475\textwidth]{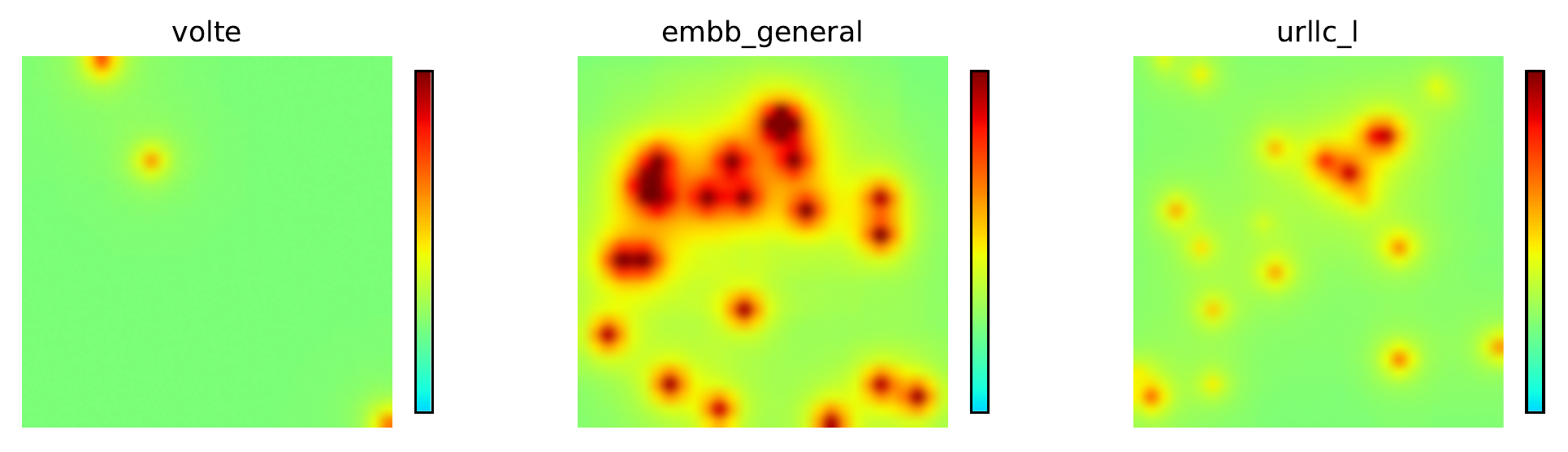}
			\label{fig:310}
		}
		\caption{The fluctuations of demands of diverse slices in the temporal and spatial domains caused by the mobility of subscribers. Different rows correspond to different iterations. The red points indicate higher mobile traffic demands while the green ones mean lower demands.}
		\label{fig:heatmap}
	\end{figure}
	
	Empirically, an outstanding resource management strategy needs to ensure the QoS for subscribers, which signifies that the successful transmission ratio of the traffic packets should be maximized as far as possible to make the network more smoothing. Thus, we involve the $ \mathrm{SSR}_{mn} $ of $n^{th}$ NS in $m^{th}$ BS, defined as the percent of successful transmitted packets in the total sent packets of $n^{th}$ NS in $m^{th}$ BS, to be one of the criteria. To calculate this variable, we use $ Q_{u_{mn}} $ to represent the set of packets sent from the $m^{th}$ BS to $u_{mn}$ which is determined by the real-time traffic demand pattern. Moreover, a zero-one variable $ x_{q_{u_{mn}}} \in \{0,1\} $ is defined as an indicator whether the transmission of packet $ q_{u_{mn}} $ conforms to the service requirement. When the SLA for service type $n^{th}$ NS in $m^{th}$ BS, downlink data rate $ \bar{r}_{mn} $ and latency $\bar{l}_{mn}$, of $u_{mn}$ are totally satisfied, $x_{q_{u_{mn}}} = 1$ means that the packet $ q_{u_{mn}} \in Q_{u_{mn}}$ is successfully received by $u_{mn}$. On the contrary, if SLA is not satisfied, $x_{q_{u_{mn}}} = 0$. Thus, $ \mathrm{SSR}_{mn} $ of $n^{th}$ NS in $m^{th}$ BS is formulated as:
	\begin{equation}
		\mathrm{SSR}_{mn} = \frac{\sum_{u_{mn} \in \mathcal{U}_{mn}} \sum_{q_{u_{mn}} \in Q_{u_{mn}}} x_{q_{u_{mn}}}}{\sum_{u_{mn} \in \mathcal{U}_{mn}} | Q_{u_{mn}} |}
	\end{equation}
	Two summation symbols in the numerator are used to sum the total successful transmission packets for all subscribers of $n^{th}$ NS in $m^{th}$ BS while the denominator are the number of whole packets of $n^{th}$ NS in $m^{th}$ BS.

	Otherwise, the traffic demands $ \bm{d}_m $ of BSs at each scheduling period are influenced by both the traffic model of each slice and the dynamic distribution of subscribers in the temporal and spatial domains. 
	To make it clear, we display the impact of subscribers' mobility on the traffic demands in Fig.~\ref{fig:heatmap}. It can be observed that as time goes by, the demands of different slices fluctuate distinctly due to the subscribers who move with different speeds. Notably, these variations of hotspots arise the frequent BS handover in the dense cellular network, which aggravates the fluctuations of service demands in the related slices and complicates the resource management in Eq.~(\ref{optimization}) in real time.
	
	For this purpose, we adopt several simplified designs of the mobility patterns\footnote{Notably, the mobility patterns could follow other models, as the proposed methods focus on learning the mobility-related fluctuations of traffic demands in each NS.} for each subscriber on the basis of straight-line motion with random bouncing (sLRB), a well-known mobility pattern defined in 3GPP \cite{3GPP2012mobility,sun2020efficient}. In particular, the trajectory and speed are fixed for each subscriber, and subscribers within the same slice have more similar mobility patterns than those in heterogeneous slices. We also assign different subscribers with various trajectories by dividing them into four groups and distributing them in the corners of the scenario with random directions and certain speeds according to the type of services. Subsequently, subscribers go forward along with certain directions until reaching the bound and then bounce following the rules of reflection. Considering the features of subscribers in various slices, the moving speed of each slice is different as lately clarified in Section V. 
	
	Therefore, besides that \cite{li2020lstm} forecasts the distribution of subscribers in time sequence, we additionally leverage GAT to reinforce the spatial cooperation among BSs in the cellular network. GAT can incorporate the states of adjacent BSs into current ones to predict the  tendency of fluctuant demands which is conducive to the resource management strategy \cite{wei2019colight}.

	
\subsection{Problem Definition}
	According to the above system model, we present the problem of RAN resource management of NS in real time as an MG. MG is one direct generalization of Markov Decision Process (MDP) that captures the mutual effect of multiple agents \cite{zhang2019multi}. Each BS in the dense cellular network is treated as an agent. Theoretically, MG is represented by a tuple
	$(\mathcal{B},\mathcal{S},\mathcal{O},\mathcal{A},\mathcal{P},\mathcal{R},\gamma)$, where $ \mathcal{B}$ denotes the set of BSs which is mentioned before. Other components of this tuple are defined as follow:
	
	a) System state space $\mathcal{S}$ and local observation space $\mathcal{O}$. In this paper, $\mathcal{S}$ denotes the system state space composed by the processed observation data from some of the agents since each agent can only obtain the local environmental data. To catch the temporal and spatial correlation of service demands, the local observation for $m^{th}$ BS at time $t$ is represented by $\bm{o}_m^t = \{\bm{d}_m^{t-1},\bm{d}_m^t\} \in \mathcal{O}$ which consists of its past and current service demands. $\bm{s}_m^t \in \mathcal{S}$ represent the system state for $m^{th}$ BS at time $t$ which is illustrated in detail in Section \ref{section_dqn}
	
	b) Action space $\mathcal{A}$. At time $t$, $m^{th}$ BS are supposed to choose an action $\bm{a}_m^t=\bm{w}_m
	$ from its candidate action space $\mathcal{A}$ as a bandwidth allocation strategy for each NS. The size of action space is determined by $\Delta$. If $\Delta$ is of coarse granularity (such as 0.54 MHz), action space will be relatively small and lead to quick convergence but the resource allocation will be not flexible enough when handling the dynamic changes of the environment and a consequently certain degree of resource waste will be involved. However, for fine granularity (such as 0.18 MHz), the action space may be too large for algorithms to converge though it can avoid the waste problem. In this paper, we simulate both coarse and fine granularity to verify the superiority of our algorithm in various conditions.
	
	c) Transition probability $\mathcal{P}$. $\mathcal{P}(\bm{s}_m^{t+1} | \bm{s}_m^t , \bm{a}_m^t)$ denotes the probability for $m^{th}$ BS to transfer from the state $ \bm{s}_m^t $ to the next state  $ \bm{s}_m^{t+1} $ according to the action $ \bm{a}_m^t $ at time $ t $.
	
	d) Reward $R$. After each time $t$, $m^{th}$ BS will obtain a real-time reward $r_m^t$ from the current environment by a specified reward function. Considering the optimization goal, the reward function is designed as:
	\begin{equation}
		\label{reward}
		r_m^t = \left\{ \begin{array}{ll}
		\displaystyle{\frac{J_m}{c_1}},&\overline {\mathrm{SSR}}_m \geqslant c_3 \vspace{2mm}\\
		\displaystyle{\frac{\overline {\mathrm{SSR}}_m}{c_2}},&\overline {\mathrm{SSR}}_m < c_3
		\end{array} \right.
	\end{equation}
	where $\overline {\mathrm{SSR}}_m$ is the average of $\mathrm{SSR}_{mn}$. $c_1,c_2$ are the constants mapping rewards to $[0,1]$ which is beneficial to the DRL training and prediction processes. $c_3$ indicates the minimum threshold of SSR to be satisfied. Such a setting is significantly different from reward clipping in \cite{hua2019gan} and reward shaping in \cite{li2020lstm}, which albeit brings performance improvement yet makes the reward functions complicated and loses the generality. Our proposed function only considers whether the bandwidth allocation strategy can guarantee the lowest SSR requirement. Once $\overline {\mathrm{SSR}}_m \geqslant c_3$, we pursue the higher $J_m$. The total accumulated return at time $t$ is $R_m^t = \sum_{k=0}^{\infty} \gamma^k r_m^{t+k}$.
	
	e) Discount factor $ \gamma $. $ \gamma \in (0,1]$ is a hyper-parameter in reward calculation which determines the importance of future rewards. Setting $\gamma=0$ implies the agent has a myopic attitude that only considers current rewards, while $\gamma=1$ attaches importance to a long-term high reward. Empirically, we set $\gamma=0.9$.

\begin{figure*}[tbp]
	\centering
	\includegraphics[width=0.8\textwidth]{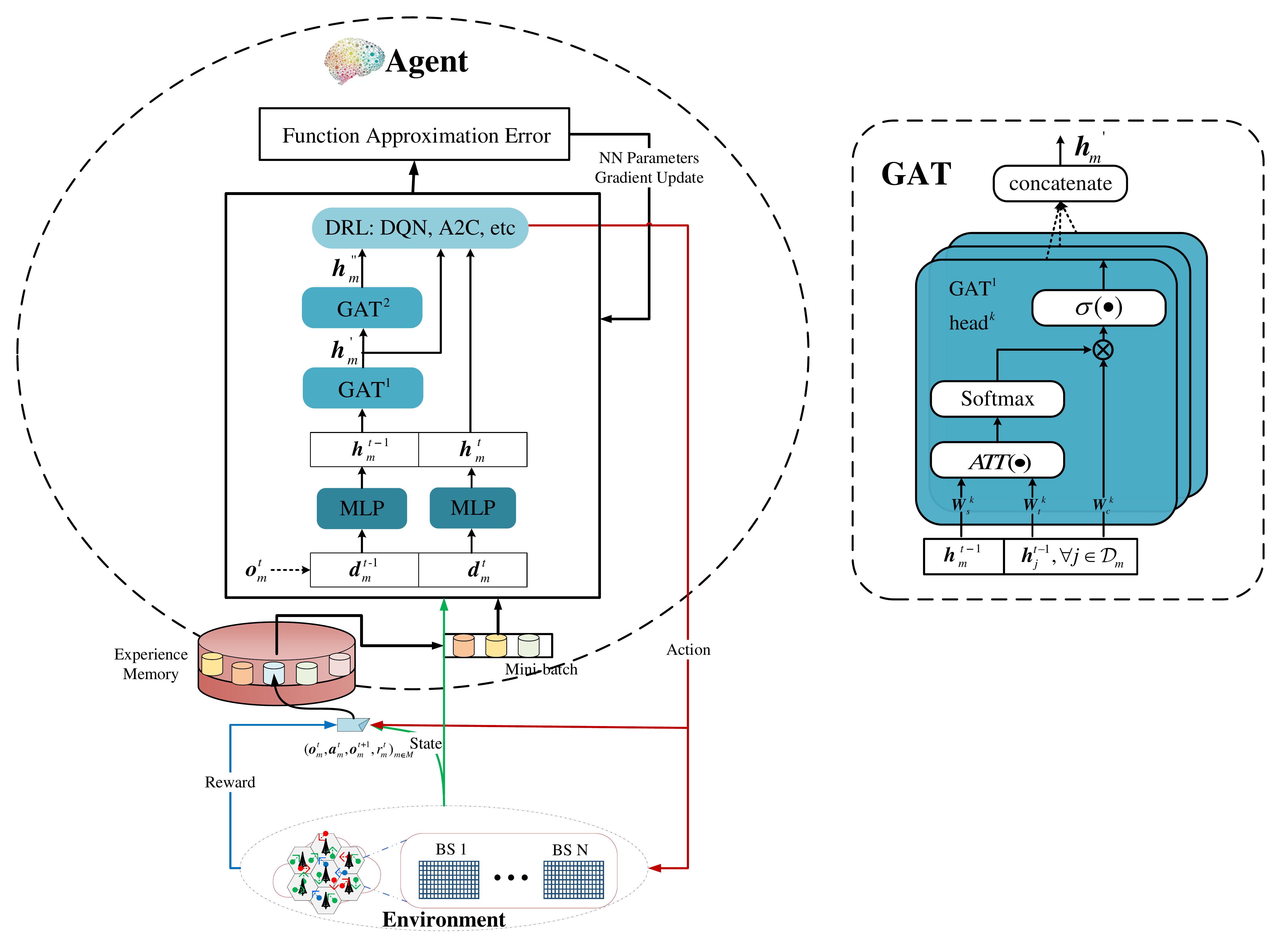}
	\caption{The illustration of GAT-based DRL algorithm for resource allocation in network slicing.}
	\label{fig:algorithm_model}
\end{figure*}

\section{The GAT-based Multi-Agent Reinforcement Learning}
	In this section, we describe the proposed GAT-based MARL algorithms, as illustrated in Fig.~\ref{fig:algorithm_model}. We introduce the network structure from bottom to top. The first step is the observation representation achieved by multi-layer perceptron (MLP), which is a non-linear function composed of a simple network including multiple layers with several neurons. Especially, due to the mobility of subscribers and the consequent BS handover, the demands of the previous step from adjacent BSs are significant features using to predict the resource management strategy in this step for the current BS. Thus, we record the past demands $ \bm{d}_m^{t-1} $ as the part of observations and process them by GAT. Notably, GAT is an effective way to process structured data which is represented as a graph. In the cellular network, the distribution of BSs can be regarded as a graph so that GAT can do the state pre-processing to track the temporal and spatial fluctuations of demands. Finally, to verify the universality of GAT for promoting the performance of DRL algorithms, we choose two classic and representative types of RL algorithms (i.e., a value-based method DQN and a combination of policy-based and value-based method A2C). We apply these dominant model-free RL algorithms to fulfill the action prediction for resource management.
	
\subsection{Observation Representation}
	For the raw data obtained from the scenario, we need to map such $ n $-dimensional vectors into a $ k $-dimensional latent space ($k > n$) by MLP for low-dimensional input or Convolutional Neural Network (CNN) for visual input, since low dimensional impartible data can be converted into high dimensional separable data by the above process. Because the raw data of our system is in low-dimension, $ \bm{o}_m^t = \{\bm{d}_m^{t-1},\bm{d}_m^t\}  $, we map it into the higher dimension by MLP, represented by:  
	\begin{equation}
		\label{MLP}
		\begin{aligned}
			\bm{h}_m^{t-1} &= MLP(\bm{d}_m^{t-1}) = \sigma(\bm{W}_e^1 \bm{d}_m^{t-1} + \bm{b}_e^1)\\
			\bm{h}_m^t &= MLP(\bm{d}_m^t) = \sigma(\bm{W}_e^2 \bm{d}_m^t + \bm{b}_e^2)
		\end{aligned}
	\end{equation}
	
	where $ \bm{d}_m^{t-1}, \bm{d}_m^t \in \mathbb{R}^n $ and $ \bm{h}_m^{t-1}, \bm{h}_m^t \in \mathbb{R}^k $. Besides, $ \bm{W}_e^1, \bm{W}_e^2 \in \mathbb{R}^{k \times n} $ and $ \bm{b}_e^1, \bm{b}_e^2 \in \mathbb{R}^k $ are the weight parameters to be trained in MLP. $\sigma$ represents the activation function which is set as ``ReLu" in this paper \cite{nair2010rectified}. Specially, the observation vector is divided into $\bm{d}_m^{t-1}$ and $\bm{d}_m^{t}$ which are treated in two MLP network separately as shown in Fig.~\ref{fig:algorithm_model}. This is due that $\bm{h}_m^{t-1}$ needs to be disposed by GAT as below while $\bm{h}_m^t$ is concatenated with the outputs of GAT and than processed by DRL.

\subsection{State Pre-processing by Graph Attention Network}
	Subscribers convert frequently among BSs which causes traffic demands fluctuating in each BS at different scheduling periods. Under this assumption, it is necessary to strengthen the cooperation among BSs which belongs to the prime issue in multi-agent reinforcement learning (MARL). If only depending on classic single-agent RL-based methods, there is no efficient way to cooperate with neighbors \cite{wei2019colight}. Hence, we achieve the purpose of state pre-processing through combining the states from adjacent BSs and computing attention coefficients between them by GAT. Referring to \cite{velivckovic2017graph,wei2019colight,jiang2018graph}, the GAT architecture is presented in the right side of Fig.~\ref{fig:algorithm_model}
	
	As the initial step, we execute the self-attention mechanism $ATT: \mathbb{R}^k \times \mathbb{R} ^ k \to \mathbb{R}$ on each BS and its adjacent BSs to calculate attention coefficients.
	\begin{equation}
		\label{self_attention}
		\begin{aligned}
		e_{mj} &= ATT(\bm{W}_s \bm{h}_m^{t-1} , \bm{W}_t \bm{h}_j^{t-1}) \\
		&= (\bm{W}_s \bm{h}_m^{t-1})^T \cdot (\bm{W}_t \bm{h}_j^{t-1})
		\end{aligned}
	\end{equation}
	where $\bm{W}_s, \bm{W}_t \in \mathbb{R}^{p \times k}$ are weight matrices to perform a shared linear transformation. This formula indicates the importance of the past state features $\bm{h}_j^{t-1}$ of $j^{th}$ BS in determining the current policy for $m^{th}$ BS.
	
	Instead of considering the effect of all other BSs for $m^{th}$ BS in the multi-agent system, we leverage graph structure into the cellular network through masked attention as in the GAT mechanism. For this, only BSs in the neighborhood will be considered when computing attention coefficients.
	\begin{equation}
		\label{mask}
		\alpha_{mj} = \mathrm{softmax} (e_{mj}) = \frac{\mathrm{exp} (\tau \cdot e_{mj})}{\sum_{j \in \mathcal{D}_m} \mathrm{exp} (\tau \cdot e_{mj})}
	\end{equation}
	where $\tau$ is the temporary factor and $ \mathcal{D}_m $ is the set of adjacent BSs including itself in the current neighborhood scope of $m^{th}$ BS defined by Euclidean distance among BSs. Besides, the ``Softmax" function is used to normalize the coefficients across different adjacent BSs in the graph to make them easily comparable.
	
	After that, these normalized attention coefficients are applied to calculate a linear combination of the states from the current BS and its neighbors to produce the output features for the current BS.
	\begin{equation}
		\label{one_att}
		\bm{h}_m' = \sigma (\sum_{j \in \mathcal{D}_m} \alpha_{mj} \bm{W}_c \bm{h}_j^{t-1})
	\end{equation}
	where $\bm{W}_c \in \mathbb{R}^{c \times k}$ is the weight matrix that needs to be trained and $ \bm{h}_m' \in \mathbb{R}^c $ is the output vector of single-head attention mechanism.
	
	Empirically, single-head attention mechanism may cause the instability of the training process of GAT, so that we extend it to multi-head attentions. It can be regarded as multiple single-head attentions executed independently in parallel while the output vectors can be concatenated or averaged. We conduct the concatenate process as follow:
	\begin{equation}
		\label{multi-head}
		\begin{aligned}
		\bm{h}_m' &=GAT(\bm{h}_m^{t-1}, \bm{h}_j^{t-1})\\
		&= \mathop {{\rm{||}}}\limits_{k = 1}^K \sigma(\sum_{j \in \mathcal{D}_m} \alpha_{mj}^k \bm{W}_c^k \bm{h}_j^{t-1} )
		\end{aligned}
	\end{equation}
	where $K$ represents the total number of multi-head attentions. 
	
	As presented in \cite{vaswani2017attention}, the more attention heads the structure has, the better relation representations and the more stable training process will be achieved. Furthermore, some researchers \cite{jiang2018graph} point out that multiple convolutional layers can extract higher order relation representations that excavate the deeper interplay and make closer cooperation between neighbors. Based on these experiences, we design the final GAT architecture for state pre-processing with two convolutional layers and eight attention heads ($K=8$) which results in the best performance. To simplify the expression, depicted in the right part of Fig.~\ref{fig:algorithm_model}, we encapsulate the formulas of Eq.~(\ref{self_attention}), (\ref{mask}), (\ref{multi-head}) for the GAT layers in the following form in which $ \bm{h}_m' $ and $ \bm{h}_m'' $ are the outputs of GAT layers respectively.
	\begin{equation}
		\label{gat}
		\begin{aligned}
		\bm{h}_m' &= \mathit{GAT}^1(\bm{h}_m^{t-1}, \bm{h}_j^{t-1}), \forall j \in \mathcal{D}_m\\
		\bm{h}_m'' &= \mathit{GAT}^2(\bm{h}_m',  \bm{h}_j'), \forall j \in \mathcal{D}_m
		\end{aligned}
	\end{equation}
	
\subsection{Resource Management by Deep Q Network and its Variants}
\label{section_dqn}
\begin{figure}[tbp]
		\centering
		\includegraphics[width=0.45\textwidth]{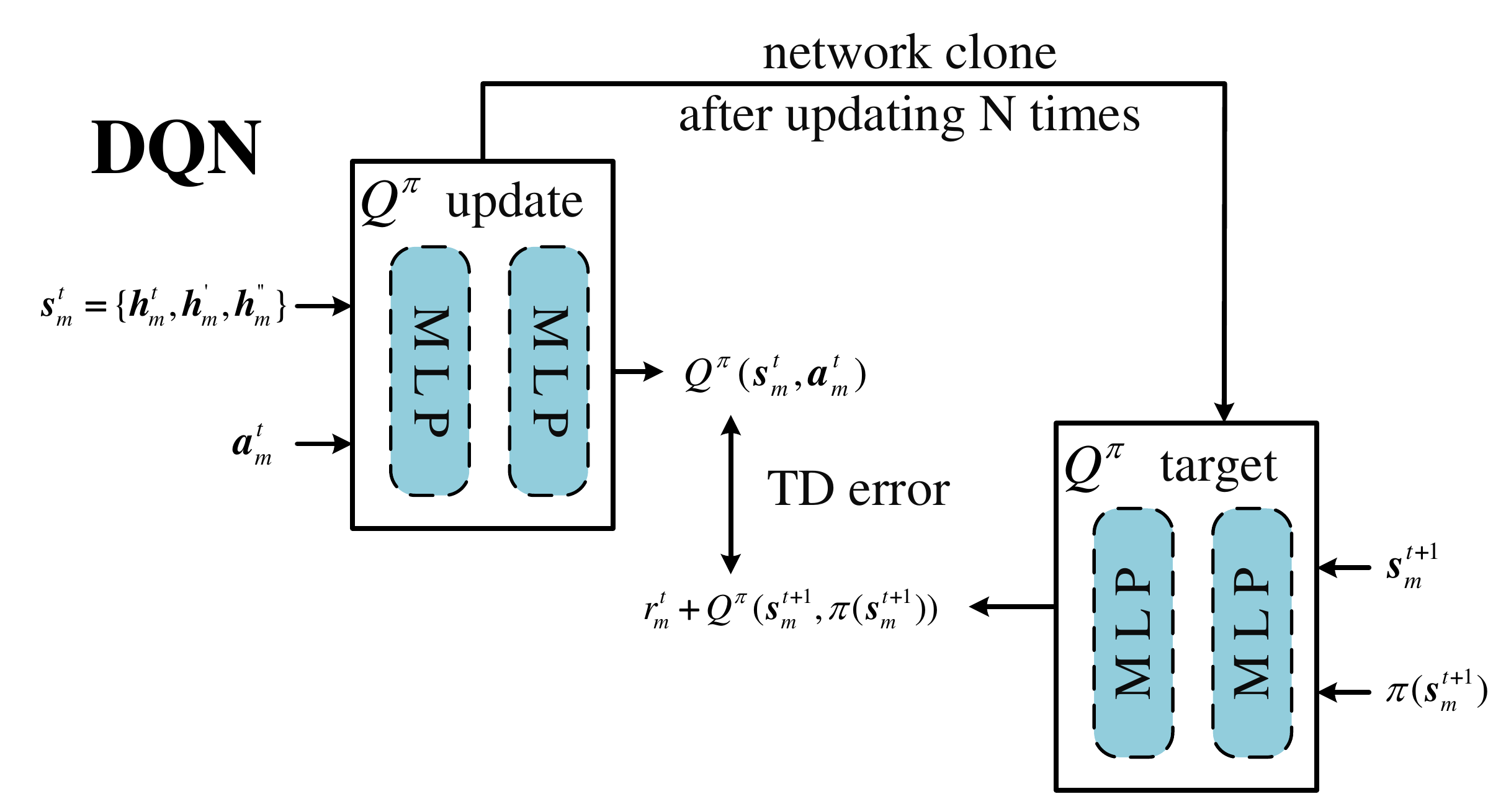}
		\caption{The illustration of resource allocation by deep Q network}
		\label{fig:dqn}
	\end{figure}

	As the final module in GAT-based DRL algorithms, we apply the standard DQN and its variants to optimizing the resource management strategy in this subsection while the details of A2C are in next subsection. DQN is based on the expectation of action-value distribution, devoting to obtaining an optimal policy $\pi(\cdot | s)$ which maps a state to a distribution over actions. According to \cite{sutton2018reinforcement,mnih2015human}, we present the training process of DQN in Fig.~\ref{fig:dqn}.

	Mathematically, the action-value function, defined as Eq.~(\ref{action-value}), denotes the expected reward of taking action $\bm{a}_m^t$ in system state $\bm{s}_m^t$ under the policy $\pi$ for agent m. 
	
	$\bm{s}_m^t=\{\bm{h}_m^t,\bm{h}_m',\bm{h}_m''\}$ is concatenated by the outputs of $MLP$ and $\mathit{GAT}^{1,2}$ .
	\begin{equation}
		\label{action-value}
		Q^{\pi}(\bm{s}_m^t,\bm{a}_m^t)=\mathbb{E}_{\pi,\mathcal{P}}[R_m^t | S=\bm{s}_m^t,A=\bm{a}_m^t]
	\end{equation}
	where $\mathbb{E}$ is the expectation. According to Bellman equation \cite{sutton2018reinforcement}, $Q^{\pi}(\bm{s}_m^t,\bm{a}_m^t)$ can be represented as:
	\begin{equation}
		\label{bellmanQ}
		Q^{\pi}(\bm{s}_m^t,\bm{a}_m^t)=\mathbb{E}_{\pi,\mathcal{P}}[r_m^t+\gamma Q^{\pi}(\bm{s}_m^{t+1},\pi(\bm{s}_m^{t+1}))]
	\end{equation}
	where $\bm{s}_m^{t+1}$ is the next system state decided by $\mathcal{P}(\cdot|\bm{s}_m^{t},\bm{a}_m^t)$.
	
	The optimal policy, pursuing the maximum $ Q^{\pi}(\bm{s}_m^t,\bm{a}_m^t) $ for all $\bm{s}_m^t$ and $\bm{a}_m^t$, is defined as:
	\begin{equation}
		\pi ^ * = \arg \max_{\pi} Q^{\pi}(\bm{s}_m^t,\bm{a}_m^t) 
	\end{equation}
	Thus, the corresponding action-value function is:
	\begin{equation}
		Q^*(\bm{s}_m^t,\bm{a}_m^t)=\mathbb{E}_{\pi^*,\mathcal{P}}[r_m^t+\gamma \max_{\bm{a} \in \mathcal{A}}Q^*(\bm{s}_m^{t+1},\bm{a})]
	\end{equation}
	
	Finally, the loss function for optimizing the current neural network is defined as:
	\begin{equation}
		\label{DQN_loss}
		\begin{array}{c}
		\displaystyle
		Y^Q=r_m^t+\gamma \max_{\bm{a} \in \mathcal{A}}Q(\bm{s}_m^{t+1},\bm{a};\theta_t)\vspace{3mm}\\
		\mathcal{L}(\theta_u) = (Y^Q-Q(\bm{s}_m^{t},\bm{a}_m^t;\theta_u))^2\\
		\end{array}
	\end{equation}
	where $ \theta_t $ and $ \theta_u $ are the target and current network trainable parameters, respectively. The target network is generated by cloning current network and updates the parameters after fixed iterations. 
	
	
	However, there exists several imperfections in the standard DQN such as overestimation and imprecision of Q value. Inspired by \cite{hessel2017rainbow}, we make several modifications to improve the performance of DQN. ``Double" and ``Dueling" are the major techniques. Double DQN \cite{van2015deep} fixes out the overestimation of Q value by decoupling, which amends the loss function as follow:
	\[
	Y^{\mathrm{doubleQ}}=r_m^t+\gamma Q(\bm{s}_m^{t+1},\arg \max_{\bm{a} \in \mathcal{A}} Q(\bm{s}_m^{t+1},\bm{a};\theta_u);\theta_t)
	\]
	\begin{equation}
		\label{ddqn}
		\mathcal{L}(\theta_u) =(Y^{\mathrm{doubleQ}}-Q(\bm{s}_m^t,\bm{a}_m^t;\theta_u))^2
	\end{equation}
	
	On the other hand, dueling network \cite{wang2016dueling} proposes two independent estimators (i.e., the value function and the action advantage function, both realized by MLP, which share the same convolutional encoder layers and calculate the values respectively while merging them in the end) to replace the single one of standard DQN and speed up the convergence. This improvement of dueling network can be presented as:
	\begin{equation}
		\begin{aligned}
		Q(\bm{s}_m^t,&\bm{a}_m^t;\theta,\mu,\nu)=V(\bm{s}_m^t;\theta,\mu)+\\
		&[ A(\bm{s}_m^t,\bm{a}_m^t;\theta,\nu)-\frac{1}{|\mathcal{A}|}\sum_{\bm{a} \in \mathcal{A}}A(\bm{s}_m^t,\bm{a};\theta,\nu) ]
		\end{aligned}
	\end{equation}
	where $\theta$, $\mu$ and $\nu$ are the trainable parameters of the shared convolutional encoder, value function $V(\cdot)$ and the action advantage function $A(\cdot)$, respectively. To sum up, we summarize the above algorithm in Algorithm 1.
	Thereinto, our algorithm uses the memory replay buffer mechanism which makes memory stay up-to-date by storing the latest sampled data and discarding the old one due to the storage constraints. At the initial phase ($t=1$ to $T/5$), where $T$ denotes the total time-step, agents interact with environment randomly to explore the state space without priori knowledge and store these samples in the replay buffer $\mathcal{F}$. After accumulating adequate samples ($t=T/5$ to $T$), the neural networks begin to be trained and updated while agents use the $\epsilon$-greedy mechanism as described in Algorithm 1 to interact with the environment and generate the sample continuously. Specially, $\epsilon$-greedy is a probabilistic selection mechanism to balance the exploration and exploitation, and determines whether the choice of agent is based on the prediction of algorithm or randomly choosing to explore the environment.
	
	\begin{algorithm}[!t]
		\caption{The GAT-based DQN algorithm}
		\label{al:DQN}
		\begin{algorithmic}[1]
			\STATE Initialize the parameters ($\theta_u \gets \text{random},\theta_t \gets \theta_u,\gamma \gets 0.9$) for the whole network composed by MLP, GAT and DQN. 
			\STATE Initialize an replay buffer $\mathcal{F} \gets \varnothing$ and the total time-step $T$;
			\STATE Set the exploration probability. $\epsilon=0$ initially and probability $p$ is sampled from $[0,1)$ at each time step for $\epsilon$-greedy.
			\FOR{$t$ = 1 to $T/5$}
			\FOR{all agent in the system}
			\STATE Obtain the current observation $\bm{o}_m^t$;
			\STATE Randomly choose and perform an action $\bm{a}_m^t\in\mathcal{A}$;
			\STATE At the end of the $t$-th scheduling period, get the next observation $\bm{o}_m^{t+1}$ and reward $r_m^t$ from environment;
			\ENDFOR
			\STATE Store transitions among all agents $(\bm{o}_m^t,\bm{a}_m^t,\bm{o}_m^{t+1},r_m^t)_{m \in M}$ in $\mathcal{F}$;
			\ENDFOR
			\FOR{$t$ = $T/5$ to $T$}
			\FOR{all agent in the system}
			\STATE Obtain the current observation $ \bm{o}_m^t $;
			\STATE Map to high dimensional through $ \bm{h}_m^{t-1} $ and $ \bm{h}_m^{t} $
			\STATE Fuse the information from neighbors by two GAT layers in Eq.~(\ref{gat})
			\STATE Use $\epsilon$-greedy to choose action and perform, $\epsilon \in [0,1)$ will be improved over time and $\bm{s}_m^t=\{\bm{h}_m^t,\bm{h}_m',\bm{h}_m''\}$:
			\[
			\bm{a}_m^t=\left\{\begin{array}{ll}
			\arg \max_{\bm{a} \in \mathcal{A}} Q^{\pi}(\bm{s}_m^t,\bm{a};\theta_u) & p \leq \epsilon\\
			\text{random} & \text{otherwise}
			\end{array} \right.
			\]
			\STATE At the end of the $t$-th scheduling period, get the next observation $\bm{o}_m^{t+1}$ and reward $r_m^t$ from environment;
			\ENDFOR
			\STATE Store transitions among all agents $(\bm{o}_m^t,\bm{a}_m^t,\bm{o}_m^{t+1},r_m^t)_{m \in M}$ in $\mathcal{F}$;
			\STATE Sample random minibatches of transitions Store transitions among all agents $(\bm{o}_m^j,\bm{a}_m^j,\bm{o}_m^{j+1},r_m^j)_{m \in M}$ from $\mathcal{F}$;
			\STATE Obtain $\bm{s}_m^j$ and $\bm{s}_m^{j+1}$ though $MLP$ and $GAT^{1,2}$ and perform a gradient descent step on Eq. (\ref{ddqn}) to update the parameters for the whole network.
			\STATE Every $ C $ steps clone $\theta_u$ to $\theta_t$
			\ENDFOR
		\end{algorithmic}
	\end{algorithm}

\subsection{Resource Allocation by Advantage Actor Critic}
\begin{figure}[tbp]
		\centering
		\includegraphics[width=0.45\textwidth]{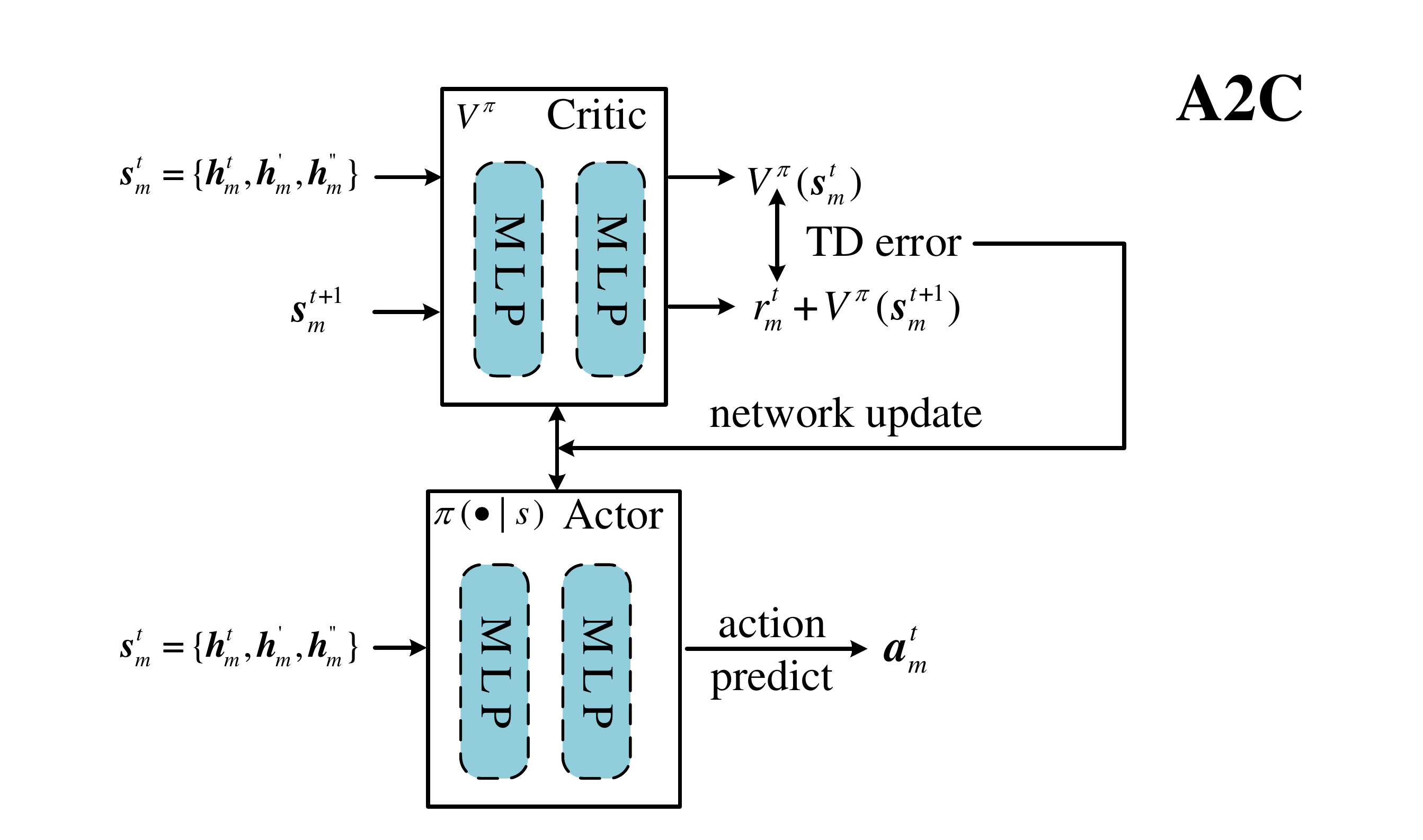}
		\caption{The illustration of resource allocation by advantage actor critic}
		\label{fig:a2c}
	\end{figure}

	Apart from DQN, we also incorporate A2C, another mainstream DRL algorithm on the basis of value-based and policy-based optimization, into GAT to demonstrate the significance of the cooperation among BSs in the multi-agent system for handling the resource management. The major steps of A2C is shown in Fig.~\ref{fig:a2c}. Unlike DQN, A2C focuses on training state-value function $V^{\pi}(\bm{s}_m^t)=\mathbb{E}_{\pi,\mathcal{P}}[R_m^t|S=\bm{s}_m^t]$ that estimates the average expected return from current state $\bm{s}_m^t$ to obtain an optimal policy $\pi(\cdot|s)$ \cite{grondman2012survey}.

	In particular, A2C is composed by two MLP networks whose inputs are similar to DQN, $\bm{s}_m^t=\{\bm{h}_m^t,\bm{h}_m',\bm{h}_m''\}$. One is ``Critic" network used to estimate state-value $V^{\pi}(\bm{s}_m^t)$. Based on mean square error (MSE) and Bellman function $V^{\pi}(\bm{s}_m^t)=\mathbb{E}_{\pi,\mathcal{P}}[r_m^t+\gamma V^{\pi}(\bm{s}_m^{t+1})]$, the loss function of this network parameters $\theta_c$ is:
	\begin{equation}
		\label{loss_critic}
		\mathcal{L}_{Critic}(\theta_c)=(r_m^t+\gamma V^{\pi}(\bm{s}_m^{t+1};\theta_c)-V^{\pi}(\bm{s}_m^t;\theta_c))^2
	\end{equation}
	
	The other is ``Actor" network which is responsible for predicting actions based on the current state. Specially, the ``Advantage" in A2C refers to $A(\bm{s}_m^t,\bm{a}_m^t)=Q^{\pi}(\bm{s}_m^t,\bm{a}_m^t)-V(\bm{s}_m^t)$ that implies the advantage of performing action $a^t$ under the state $\bm{s}_m^t$ \cite{li2020lstm}. To simplify the network structure, we apply some transformations that
	\begin{equation}
		\begin{aligned}
		A(\bm{s}_m^t,\bm{a}_m^t)&=Q^{\pi}(\bm{s}_m^t,\bm{a}_m^t)-V(\bm{s}_m^t)\\
		&\thickapprox r_m^t+\gamma V^{\pi}(\bm{s}_m^{t+1}|\bm{s}_m^t,\bm{a}_m^t)-V^{\pi}(\bm{s}_m^t)\\
		&=\delta (\bm{s}_m^t)
		\end{aligned}
	\end{equation}
	which is the Temporal-Difference (TD) error \cite{sutton2018reinforcement} of ``Critic" network. To obtain an optimal policy that executes the most valuable action under current state, this ``Advantage" is involved in the loss function of ``Actor" network parameters $\theta_a$ as \cite{li2020lstm}:
	\begin{equation}
		\begin{aligned}
		\mathcal{L}_{Actor}(\theta_a) =& -[\delta(\bm{s}_m^t;\theta_c)\log \pi (\bm{a}_m^t|\bm{s}_m^t;\theta_a)\\
		&+\lambda H(\pi (\bm{a}_m^t|\bm{s}_m^t;\theta_a))]
		\end{aligned}
		\label{loss_actor}
	\end{equation}
	where entropy regularization $H(\cdot)$ is used to encourage exploration in large action space and forbid the algorithm from converging to local optimum. $\lambda$ is the weight parameter for regularization.
	
	The algorithm of GAT-A2C is similar to GAT-DQN in Algorithm \ref{al:DQN}, thus only some special details are pointed out: 
	\begin{itemize}
		\item In the training process, we sample random minibatches of transitions $(\bm{o}_m^j,\bm{a}_m^j,\bm{o}_m^{j+1},r_m^j)_{m \in M}$ from $\mathcal{F}$ to train the ``Critic" network and obtain the TD error of state-value functions. TD error is used to perform a gradient descent step on Eq. (\ref{loss_critic}) and (\ref{loss_actor}) to update the parameters of ``Critic" and ``Actor" network, respectively.
		
		\item In the predicting process, agent, $m^{th}$ BS, selects the action $\bm{a}_m^t$ based on $\bm{s}_m^t$ depending on ``Actor" networks $\pi (\bm{a}_m^t|\bm{s}_m^t)$.
		
		\item In the location, our A2C algorithm is integrated in the agents (BSs). Each agent plays the ``Critic" and ``Actor" simultaneously while different agents use the independent networks and cooperate with others by GAT.
	\end{itemize}

\section{Simulation results and numerical analysis}
\subsection{Simulation Environment Settings}

	\begin{table}[tbp]
		\centering
		\caption{A Summary of Key Settings for Traffic Generation Per Slice}
		\label{tb:configuration}
		\begin{tabular}{|m{20pt}|m{30pt}|m{30pt}|m{53pt}|m{35pt}|}
			\toprule
			\multicolumn{2}{|c|}{} & VoLTE & eMBB & URLLC \\ \midrule
			\multicolumn{2}{|c|}{Bandwidth/$\Delta$} & \multicolumn{3}{c|}{10 MHz/0.18 MHz/0.54 MHz} \\ \hline
			\multicolumn{2}{|c|}{Scheduling}  & \multicolumn{3}{c|}{Round robin per slot (0.5 ms)}  \\ \hline
			\multicolumn{2}{|c|}{Slice Band Adjustment}  & \multicolumn{3}{c|}{1 second (2000 scheduling slots)} \\ \hline
			\multicolumn{2}{|c|}{Channel}  & \multicolumn{3}{c|}{Rayleigh fading}  \\ \hline
			\multicolumn{2}{|c|}{Base Station No} & \multicolumn{3}{c|}{19} \\ \hline
			\multicolumn{2}{|c|}{Subscriber No.(2000 in all)}& 333 & 667 & 1000 \\ \hline 
			Speed & Varying  & Uniform [Min: 1m/s, Max: 5m/s] & Uniform [Min: 1m/s, Max: 3m/s]  & Uniform [Min: 6m/s, Max: 10m/s] \\ \hline
			\multicolumn{2}{|p{70pt}|}{Distribution of Inter-Arrival Time per Subscriber} & Uniform [Min: 0ms, Max: 160ms] & Truncated  Pareto [Exponential Para: 1.2, Mean: 6ms, Max: 12.5 ms]  & Exponential [Mean: 180ms]  \\ \hline
			\multicolumn{2}{|p{70pt}|}{Distribution of Packet Size} & Constant: $40$  Byte & Truncated Pareto [Exponential Para: 1.2, Mean: 100 Byte, Max: 250 Byte] & Variable Constant: $\{$0.3, 0.4, 0.5, 0.6, 0.7$\}$ MByte  \\ \hline
			\multicolumn{1}{|l|}{\multirow{2}{*}{SLA}} & \multicolumn{1}{l|}{Rate} & 51kbps & 100 Mbps & 10 Mbps \\ \cline{2-5} 
			\multicolumn{1}{|l|}{}  & \multicolumn{1}{l|}{Latency} & 10 ms  & 10 ms & 3 ms \\ \bottomrule
		\end{tabular}
	\end{table}

	\begin{figure*}[h]
		\centering
		\subfigure{
			\centering
			\includegraphics[width=0.495\linewidth]{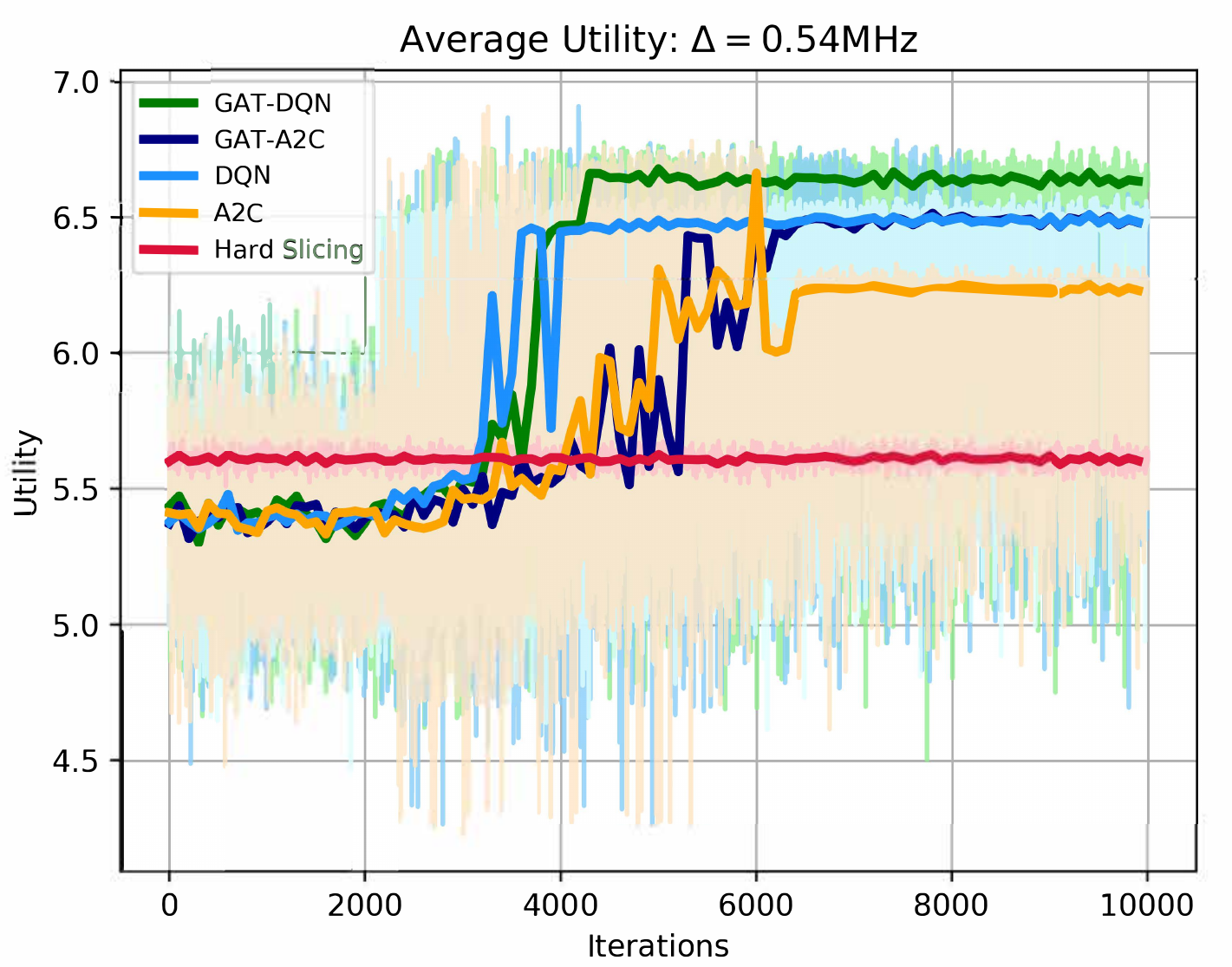}
			\label{fig:utility054}
		}\subfigure{
			\centering
			\includegraphics[width=0.495\linewidth]{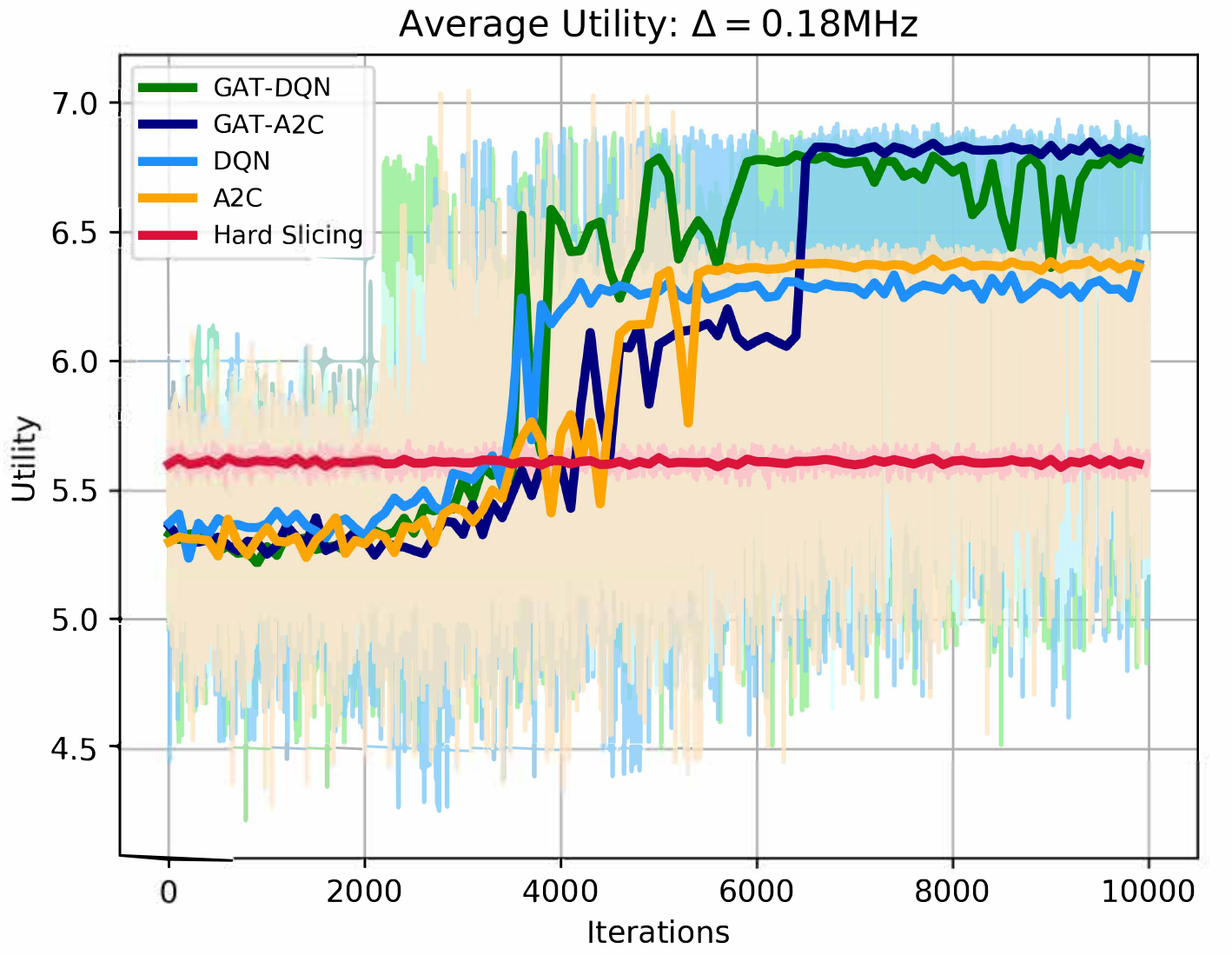}
			\label{fig:utility018}
		}
		\caption{A performance comparison of the system utility under two optional granularity between different algorithms. The shadow of each color implies the true average value of all BSs in each iteration while the curve with the corresponding color is composed by the median values for every 50 iterations. Because the true value sequences contain some values of random exploration which is meaningless, these median curves can ignore the influences caused by these values so as to be more visualized than the true value sequences.}
		\label{fig:utility}
	\end{figure*}
	\begin{figure}[tbp]
		\centering
		\includegraphics[width=0.495\textwidth]{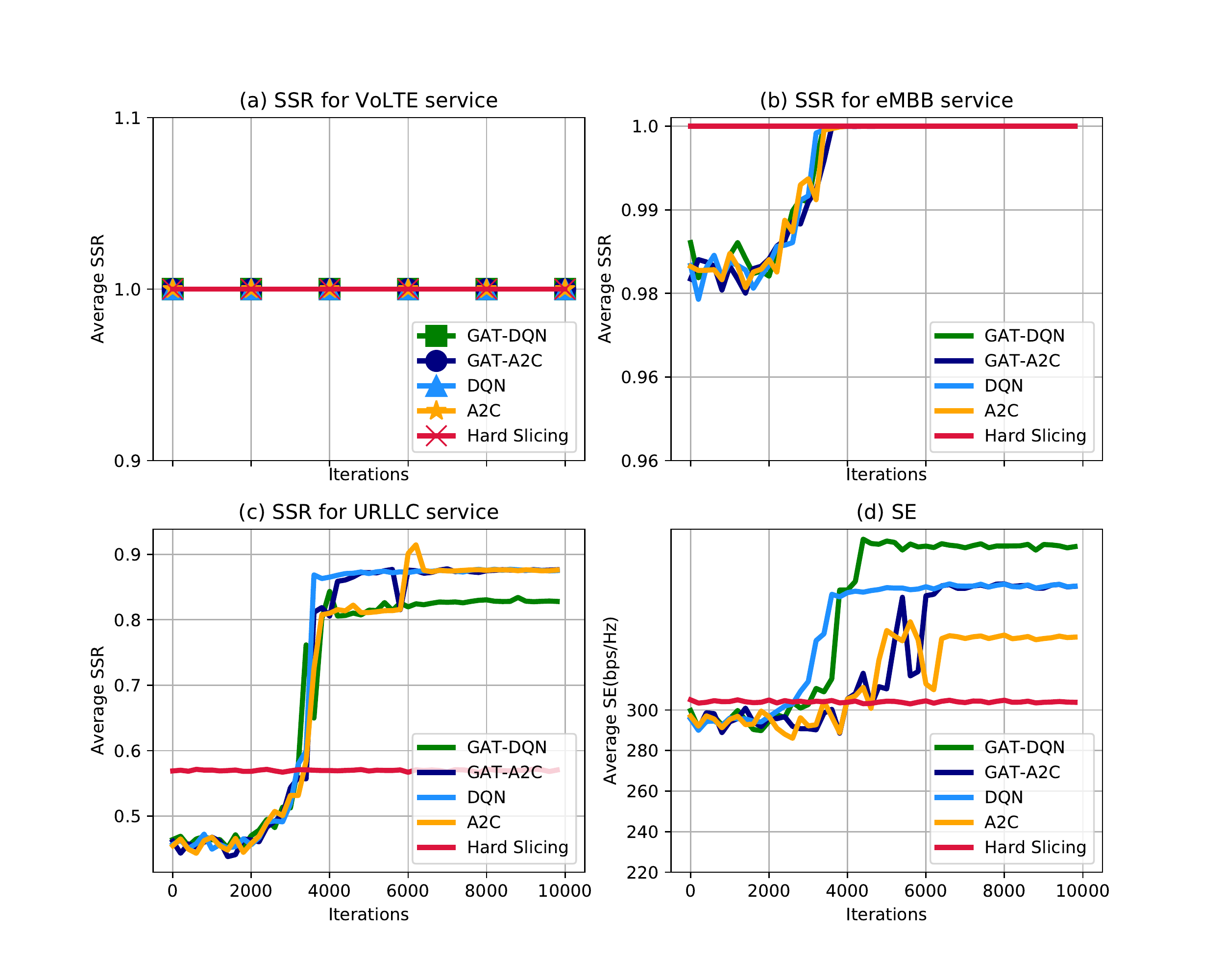}
		\caption{Several detail indicators (i.e., SSR for each service and total SE) of the system utility in granularity of $\Delta=0.54$ MHz.}
		\label{fig:qoe054}
	\end{figure}
	
	\begin{figure}[tbp]
		\centering
		\includegraphics[width=0.495\textwidth]{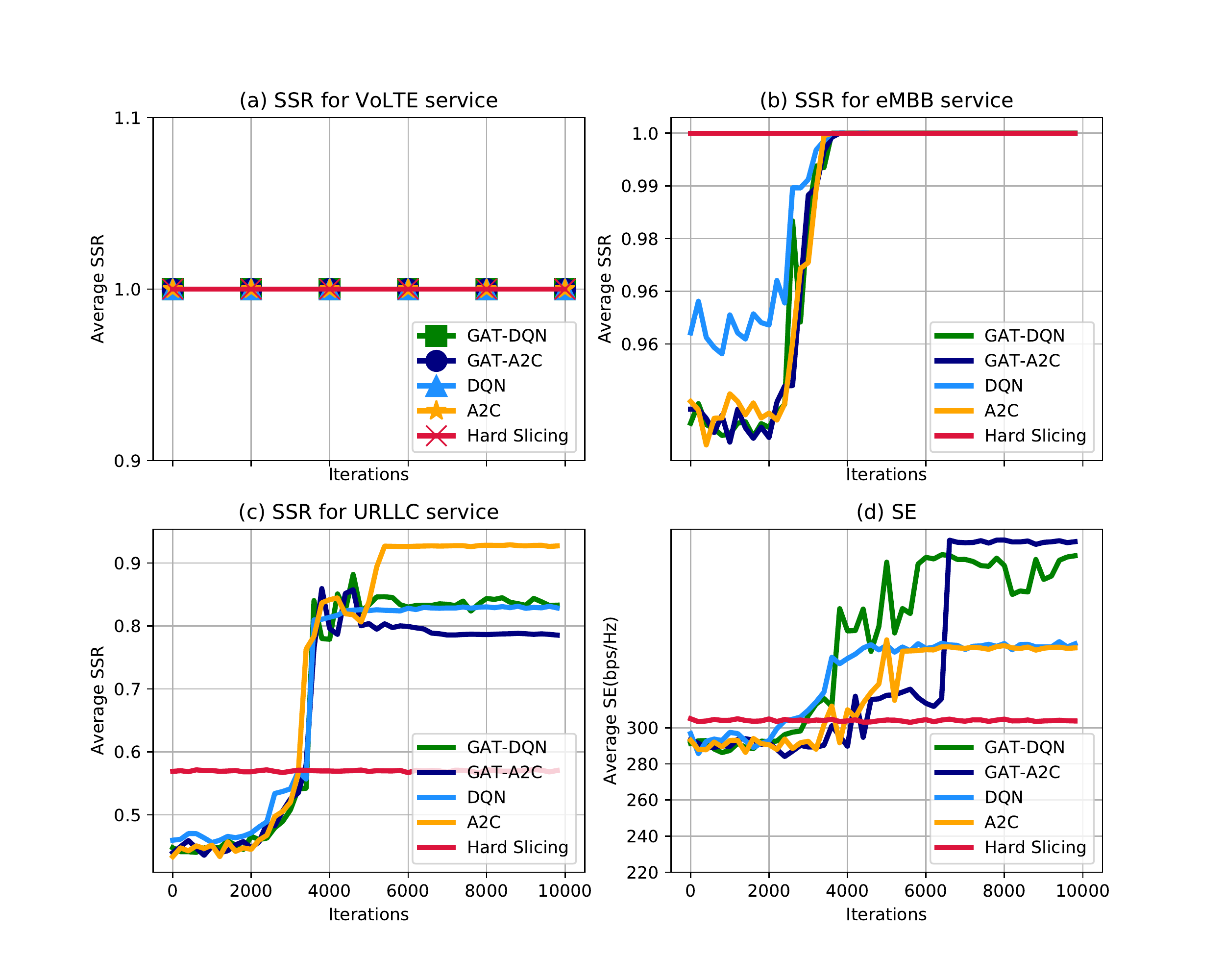}
		\caption{Several detail indicators (i.e., SSR for each service and total SE) of the system utility in granularity of $\Delta=0.18$ MHz.}
		\label{fig:qoe018}
	\end{figure}
	Based on the aforementioned multi-agent scenario, we consider $ 19 $ BSs arranged like beehives as displayed in Fig.~\ref{fig:system_model} to simulate a dense cellular network environment which is 160m $\times$ 160m in size and contains 2000 subscribers. The total bandwidth is $ 10 $ MHz with two optional granularity (i.e., 0.54 MHz for coarse granularity and 0.18 MHz for fine granularity which are the multiples of resource blocks.) in this section. For simplicity, our simulation only involves three typical services in daily life with diverse SLA (i.e., VoLTE for voice communication, eMBB for HD video transmission, and URLLC for industrial-grade application) for each BS to conduct the independent inter-slice resource management. The service demands produced by subscribers are briefly summarized in Table \ref{tb:configuration} referring to 3GPP TR 36.814 and TS 22.261 \cite{3GPP2,3GPP3}. Every 1 second, we reallocate the bandwidth to each slice to achieve real-time resource isolation and sharing between slices, which contributes to ensuring the QoS and improves the resource utilization. Within each second, each slice re-allocates its bandwidth to each subscriber every 0.5 millisecond according to the specific rules (round-robin scheduling in this paper) of the slice. In both coarse and fine granularity, we set $c_1=6, c_2=2$ in Eq. (\ref{reward}) of the hyper-parameters for reward definition. Moreover, we simulate the URLLC service with relatively large size packets as shown in Table \ref{tb:configuration} so we set a moderate threshold of reward definition $c_3=0.9$.

\subsection{Simulation Results}
	To show the significance of state pre-processing in strengthening the cooperation among BSs by GAT, we incorporate two aforementioned DRL algorithms (DQN with its variants, as well as A2C) to GAT and conduct the simulations under the above environment settings. DRL-based schemes (DQN and A2C) and hard slicing methods are involved as baselines to make the performance improvement more obvious. Hard slicing allocates the total bandwidth for each slice uniformly in which one of them can obtain $\frac{1}{N}$ of the bandwidth (there are three types of services in total thus $N=3$). Additionally, the baselines of DRL-based resource management schemes in this paper are similar to the proposed algorithms except for having no GAT structure. Due to the setting of channel mode in our simulation, the value of SE is on the scale of hundreds while the value of SSR is within $[0,1]$. Considering the magnitude of SE and SSR, the hyper-parameters of weighted sum in the optimization function Eq. (\ref{optimization}) are set to $\alpha=0.01$ and $\bm{\beta}=[1,1,1]$. 
	
	Fig.~\ref{fig:utility} depicts the performance comparison of system utility between different algorithms  under the two optional granularity. The two different granularity simulations aim to demonstrate and verify the convergence of algorithms under different sizes of action space. The left part of Fig.~\ref{fig:utility} depicts the variations of system utility with respect to the iterations under the coarse granularity, $\Delta=0.54$ MHz, which provides smaller action space. Obviously, these DRL-based algorithms achieve satisfactory performance improvements in system utility after several training steps compared with hard slicing. Although all DRL-based algorithms converge finally, these Q-learning algorithms (DQN and GAT-DQN) converge faster than Actor-Critic algorithms (A2C and GAT-A2C) (Q-learning algorithms are stable after $ 4000^{th}$ iteration while it takes Actor-Critic algorithms near $ 6000^{th}$ iteration to converge). Meanwhile, Q-learning algorithms perform better utility slightly in small action-space ($\Delta=0.54$) than Actor-Critic algorithms while they are less well in larger action-space ($\Delta=0.18$) especially in terms of stability after convergence. Notably, the GAT mechanism promotes the agents to find a superior policy resulting in an improvement for GAT-DQN and GAT-A2C algorithms compared with DQN and A2C. The left part of Fig.~\ref{fig:utility} indicates that the result of GAT-DQN is around $ 6.8 $, which is $ 4 $ percent higher than DQN while the result of GAT-A2C increases almost $ 5 $ percent. The same conclusion can be drawn from the right part, for the fine granularity, $\Delta=0.18$ MHz, which results in larger action space. GAT-DQN and GAT-A2C have the similar performance which almost reaches the utility in $ 6.8 $ and increases $ 7 $ percent than DQN and A2C while GAT-A2C yields a more stable converging curve. In this regard, our algorithms address the shortage of vanilla DRL-based algorithms which easily result in a suboptimal solution regardless of the size of action space.
	
	\begin{figure}[htbp]
		\centering
		\includegraphics[width=0.495\textwidth]{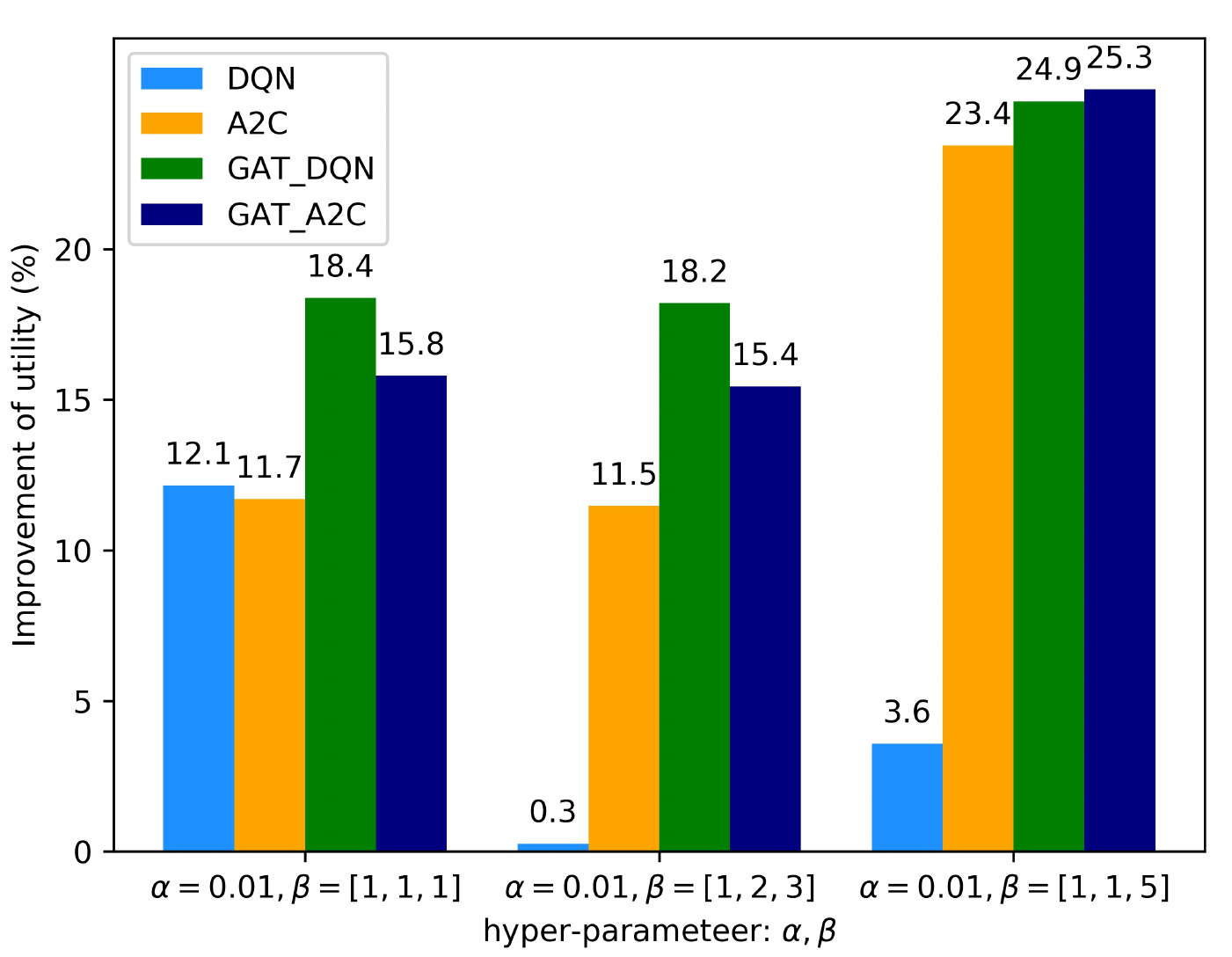}
		\caption{Performance comparison among different hyper-parameters. X-axis represents different parameter combinations and Y-axis means the improvement of utility based on the hard slicing algorithm. Different color bars represents the utility of different algorithms illustrated in the legend.}
		\label{fig:parameter_compare}
	\end{figure}
	
	In addition, we provide several detail indicators (SE and SSR for each slice) that are the compositions of system utility as shown in Fig.~\ref{fig:qoe054} for the $\Delta=0.54$ MHz case and Fig.~\ref{fig:qoe018} for the $\Delta=0.18$ MHz case. It can be observed that with respect to SSR, all DRL-based algorithms bring significant improvement to satisfy the SLA for URLLC subscribers (between 0.8 and 0.9) while not decreasing the SLA of other subscribers (almost 1.0). On the other hand, all DRL-based algorithms also increase the SE compared with hard slicing. In both $\Delta=0.54$ MHz and $\Delta=0.18$ MHz cases, with the help of GAT, the actions predicted by GAT-DQN and GAT-A2C can give a higher SE on the condition of ensuring the same SSR. Although GAT-DQN algorithm in $\Delta=0.18$ MHz case and GAT-A2C algorithm in $\Delta=0.18$ MHz case perform a slightly inferior in SSR for URLLC service than others, they reach the outstanding results in SE. This is due to the setting of reward function that once the mean SSR reaches the specified value (0.9 in this version), it will pursue higher SE performance.
	
	Besides, we measure the performance of different algorithms under diverse combinations of hyper-parameters. Considering the scale of SE, we fix $\alpha=0.01$ and change the $\bm{\beta}$ to adjust the influence from different slices. In this part, we choose three values of $\bm{\beta}$, $\bm{\beta}=[1,1,1], [1,2,3], [1,1,5]$ and the related parameter $c_1=6,9,10$ is changed to fit the optimization function. We present the comparison chart in Fig.~\ref{fig:parameter_compare}. This chart shows the utility improvement compared with hard slicing under 0.18 MHz for coarse granularity. It presents that no matter how the parameters are set, RL can always improve performance with little manual adjustment while GAT is icing on the cake.
	
	\textbf{\textit{Remark}}: There are several conclusions that we sum up from these simulation results: (a) GAT mechanism can improve the utility performance through enhancing the cooperation among individual BSs; (b) GAT-based DRL algorithms are predominant regardless of the size of action space while this advantage is more significant in the large action space; (c) These algorithms powered by Q-Learning present better results of convergence speed while Actor-Critic based algorithms perform better in terms of stability after convergence.

\section{Conclusion}
	In this paper, we have proposed to use GAT to strengthen the cooperation among BSs in the dense cellular network to capture the patterns of fluctuant service demands in temporal and spatial, and combined it with mainstream DRL algorithms to yield an intelligent resource management strategy for NS. For verifying the universality of GAT in promoting the performance of DRL algorithms, we have selected two classic and representative algorithms of DRL (i.e., DQN and its variants, as well as A2C). Extensive simulation results have demonstrated that incorporating GAT for state pre-processing on the top of these DRL algorithms is effective to enhance the cooperation and obtain the optimal policy for the multi-BS system in RAN. It can not only satisfy the strict SLA requirements but also improve the SE indicator, thus providing a promising solution in slicing resource management. Nevertheless, many subsequent issues need to be addressed in the future, such as the verification of its robustness facing more severe environment in reality, the demonstration of its capability to deal with interference and complex mobility pattern, the improvement of neural network structure to reduce the computational complexity such as COMA \cite{Foerster2018COMA}, the comprehensive comparison with the existing algorithms in resource management.
	
\bibliographystyle{IEEEtran}
\bibliography{references}

\end{document}